\documentclass[final,5p,times,twocolumn]{elsarticle}

\usepackage{amssymb}

\usepackage{amsmath}

\newcommand{\C}{$\,^{\circ}\mathrm{C}$}

\journal{Current Opinion in Solid State \& Materials Science}

\begin{document}

\begin{frontmatter}



\title{First-principles derived complexion diagrams for phase boundaries in doped cemented carbides}


\author{Sven A. E. Johansson}
\author{G. Wahnstr{\"o}m \corref{cor1}}
\ead{goran.wahnstrom@chalmers.se}
\cortext[cor1]{Corresponding author}

\address{Department of Applied Physics, Chalmers University of Technology, SE-412 96 G{\"o}teborg, Sweden}

\begin{abstract}
This article reviews a method for calculating an equilibrium interfacial phase diagram depicting regions of stability for different interface structures as function of temperature and chemical potentials. Density functional theory (DFT) is used for interfacial energies, Monte Carlo simulations together with cluster expansions based on DFT results for obtaining configurational free energies, and CALPHAD-type modeling for describing the thermodynamic properties of the adjoining bulk phases. An explicit case, vanadium doped cemented carbides, is chosen to illustrate the progress in the research field and the interfacial diagram, a complexion diagram, for the phase boundary WC(0001)/Co is constructed as function of temperature and chemical potentials.
\end{abstract}


\end{frontmatter}


\section{Introduction}

Computational techniques have been developed to model properties of complex multicomponent materials. Methods such as CALPHAD are used to describe the thermodynamic phase behavior \cite{LuFrSu07}. Experimental data and in some cases results from first-principles methods are then utilized in an efficient way to compute the corresponding phase diagrams.

For a polycrystalline material, not only the phase behavior is important, but also the structure, composition and energetics of its internal interfaces are decisive for the properties of the material. The study of interfaces is hence of great importance and is a growing and rapidly advancing field within materials science. It has been well established that the appearance of certain interfacial ``phases'' different from those of the adjoining bulk phases (e.~g. segregation structures, amorphous wetting films, etc.) can be understood from thermodynamic considerations \cite{CaTaDi14,Lu07,JeWaMi09,JoWa10,Ha11,BaChKa11,JoWa11,Lu12,JoWa12,FrOl13,GaWi14,TeNaPa15,RhHo15,KhRu15}; the disparate interfacial phases are equilibrium states stabilized by the interface. The term {\it complexion} has been suggested to separate these interfacial phases from ordinary bulk phases \cite{TaCaCa06,DiTaCaHa07,DiHaLu09}.

Development of computational methods to derive the corresponding {\it interfacial phase diagrams} has therefore become important. In contrast to the bulk behavior experimental data for the interfacial properties are scarce and hence one may expect results from first-principles methods to become more critical and decisive in the development of interfacial phase diagrams compared with ordinary bulk phase diagrams.

Here we review how interfacial phase diagrams can be constructed by combining results from first-principles density functional theory (DFT) calculations together with thermodynamic modeling of the adjoining bulk phases. An explicit case, vanadium doped cemented carbides, is chosen to illustrate the progress in the research field \cite{JoWa10,JoWa11,JoWa12}.

\section{Cemented carbides}

Cemented carbides is a class of materials which consists of fine particles of a carbide cemented into a composite by a binder metal. The most common cemented carbide is produced by combining tungsten carbide (WC) with the binder metal cobalt (Co). Cemented carbides combine excellent hardness with high toughness and wear resistance \cite{ComprHardMater14}. They are used in cutting and wear resistant tools and are of considerable industrial importance \cite{NoGaBl15}.

The material is produced in a sintering process, in which carbide and metal powders densify into a microstructure of faceted WC grains embedded in the Co binder. Considerable research work has been focused on producing cemented carbides with WC grain sizes in the nanometer scale, since such a fine microstructure has the potential to dramatically improve on the mechanical properties of the material \cite{FaWaRyHwSo09}. To mitigate WC grain growth and retain the fine structure of the starting powder, additions of VC, $\text{Cr}_3\text{C}_2$, NbC, TiC, TaC, etc. are made, of which especially VC works effectively as grain growth inhibitor \cite{ScBoLu95}. The effects on microstructure of the additions have to large extent been empirically investigated, but still the exact mechanism behind the grain growth inhibition is unknown.

Several high-resolution electron microscopy studies have shown WC/Co interfaces that contain films with a thickness of a few atomic layers having an atomic arrangement incompatible with the underlying hexagonal WC lattice. The films have been seen especially in VC-doped WC-Co \cite{JaYaIkSaTaOkTa98,YaIkSa00,LaHaThLa02,LaThHaTh03,LaLoDo04,LaHaThLo04,LaLoJoWa12,SuMiTa12,SuMiTa13,SuMiTa15}, where they have been associated with the cubic VC phase \cite{LaHaThLa02}. Thin films in WC/Co interfaces have also been imaged in $\text{Cr}_3\text{C}_2$- \cite{DeBaPaLaAl04} and TiC-doped \cite{KaTeHa06} cemented carbides.

It has been suggested, that thin films in WC/Co interfaces hinder the WC grain growth by forming a diffusion barrier for dissolving and/or reprecipitating W atoms \cite{LaThHaTh03}. However, the observed thin films at the WC/Co interface in the final material does not necessarily correspond to the equilibrium state during sintering \cite{KaTeHa04}. The sintering process occurs at high temperature -- including temperatures above the melting of the Co binder phase -- where solubilities of W, C and dopant atoms in Co are high and where the interface region can be expected to be in a state of local chemical equilibrium. During cooling in the end of the sintering cycle, most of the dissolved atoms leave the Co binder phase \cite{HaAg98} and reprecipitate onto existing carbide grains. It is thus conceivable that experimentally observed atomic structures and dopant segregation correspond to a low-temperature equilibrium or some frozen-in state from the end of the cooling
\cite{KaKi15a,KaKi15b}. However, experimental studies using rapid temperature quench of the samples give strong evidence for the existence of thin films in WC/Co interfaces at the holding temperature during liquid phase sintering \cite{SuMiTa12}.

As most of the microstructural development occurs at high temperatures, it is essential to know the equilibrium state of the WC/Co interface in a doped WC-Co system as function of temperature and dopant addition {\it i.e.} one has to know the corresponding {\it interfacial phase diagram}. Only with this knowledge can a connection between WC growth kinetics and WC/Co interfacial structure and segregation eventually be established.

\section{Simplified modeling}

WC has a hexagonal lattice structure in cemented carbides. In the phase diagram, WC exists in a very small range of homogeneity with a carbon content corresponding to nearly perfect stoichiometry, meaning that in practice W and C vacancies can be assumed not to occur \cite{MaSuFr05}. Furthermore, the solubility of transition metal atoms in WC is small. For V it is found to be only about 10$^{-3}$ atomic fractions at 1410\C\ \cite{WeJoAnWa11}.

The hexagonal WC structure is the stable structure over the whole temperature interval relevant for sintering (below 2000\C). However, a cubic WC phase can be obtained at sufficiently high temperatures (above 2500\C) \cite{Gu86}. In contrast, the stable structures for other carbides, such as VC, TiC, and TaC, are cubic and the stoichiometry can deviate substantially from perfect.

During sintering, the WC grains develop a faceted triangular shape, bounded by basal \{0001\} and prismatic \{10$\bar{1}$0\} planes. Theoretical data for the WC/Co interface energies for the basal and prismatic planes are in the range 0.9-1.5 J/m$^2$ \cite{PeJoWa15}. For the cubic WC structure the WC/Co interface energies are lower, about 0.5 J/m$^2$ for the WC(100)/Co interface \cite{ChDuWa02}.

In cemented carbides the stable lattice structure for WC is the hexagonal one. However, due to the lower WC/Co interface energy for cubic WC compared with hexagonal WC, one could speculate that the atomic stacking of the carbide near a WC/Co interface could transform locally into cubic to lower the interface energy. Indeed, experimentally a few layers of cubic stacking has been identified at some WC/Co interfaces in undoped WC-Co \cite{BoLaLoMi08}.

An intriguing scenario is then that in doped cemented carbides at low doping conditions the dopants may preferentially segregate to WC/Co interfaces and locally form intermixed cubic structures at the interface, {\it interface complexions}, stabilized by interfacial effects. At higher doping conditions the corresponding cubic carbide phase would precipitate, which may give undesirable properties to the cemented carbide material. With carefully controlled doping it would then be possible to preferentially influence the properties of the interfaces in the cemented carbide, while limiting the precipitation of the cubic carbide phase.

\subsection{Simplified modeling}

\begin{figure}[h]
	\includegraphics[width=\columnwidth]{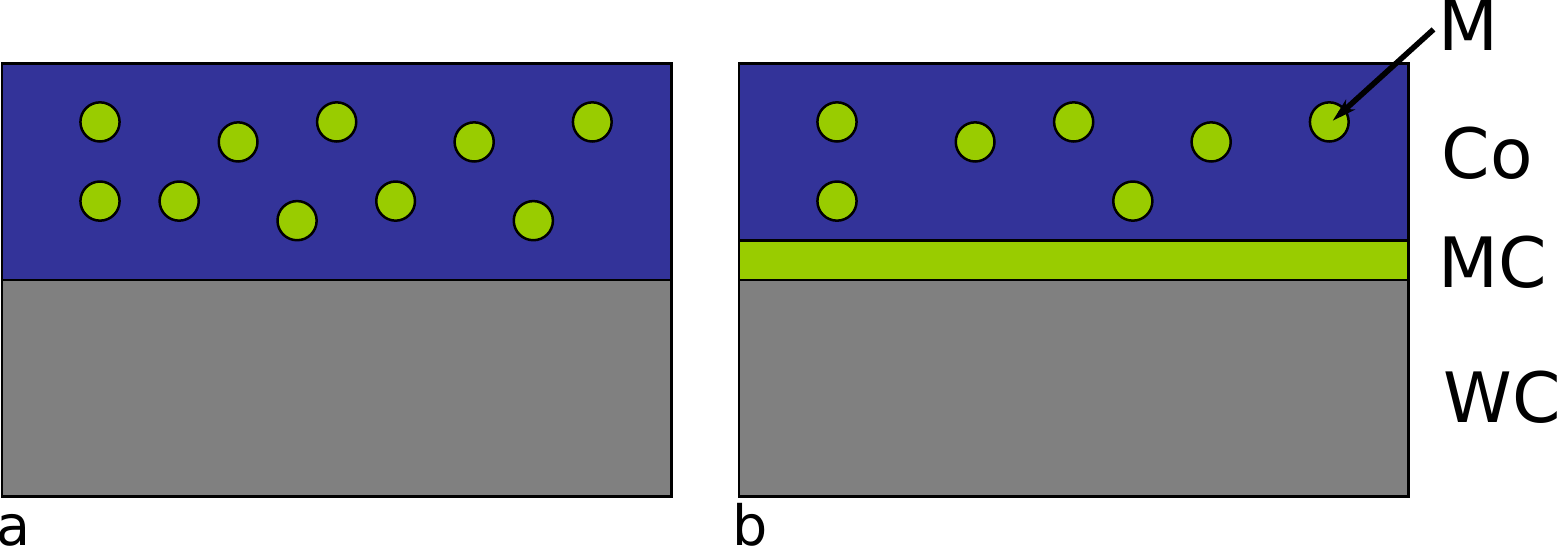}
	\caption{\label{fig:simple_model}Illustration of the simplified model. (a) A WC/Co interface with metal atoms M dissolved in the Co binder phase. (b) A thin MC film is formed at the WC/Co interface.}
\end{figure}

To model the possibility to create intermixed cubic structures, interface complexions, at WC/Co interfaces in doped cemented carbides we start with a very simple model. Consider a sharp interface between WC and the binder phase. The latter consists of mainly Co but also some dissolved metal atoms M, such as V, Ti or Ta, the dopants (see Fig.~\ref{fig:simple_model}(a)). We then assume that the interface is replaced by a thin film of a cubic MC phase (see Fig.~\ref{fig:simple_model}(b)). For a sufficiently thick film the interface can be regarded as split into a WC/MC interface and a MC/Co interface separated by a slab of MC. The energy for such a film-covered interface can then be decomposed as
\begin{equation}
	\label{eq:gamma_film}
	\gamma_\text{film} = \gamma_\text{WC/MC} + \gamma_\text{MC/Co} + N \left(  \Delta g_\text{MC} +  e_\text{MC} \right),
\end{equation}
where $\gamma_\text{WC/MC}$ ($\gamma_\text{MC/Co}$) is the interface energy between WC and MC (MC and Co). The last term of Eq.~(\ref{eq:gamma_film}) represents the energetic cost per layer of building the MC phase and, thus, depends linearly on the number of layers $N$ of the film. The energy $\Delta g_\text{MC}$ is the negative of the driving force of nucleation of the bulk MC phase. It is positive if the MC bulk phase is unstable, if we are below the solubility limit of the metal atoms M in the binder phase. We consider coherently strained thin films and the last quantity of Eq.~(\ref{eq:gamma_film}), $e_\text{MC}$, is the strain energy per layer needed to bring the film into epitaxy with the underlying hexagonal WC phase.

The energy $\gamma_\text{film}$ should be compared with the energy $\gamma_\text{WC/Co}$ of a WC/Co interface not covered by any MC film. The film coveraged interface is stable if $\gamma_\text{film} < \gamma_\text{WC/Co}$ and, hence, a necessary condition for film formation in this simplified model is that
\begin{equation}
	\label{eq:deltaGamma}
	\Delta \gamma_\text{MC} \equiv
    \gamma_\text{WC/Co} - \left( \gamma_\text{WC/MC} + \gamma_\text{MC/Co} \right) > 0
\end{equation}
{\it i.e.} one has to gain interfacial energy when the two interfaces $\text{WC/MC}$ and $\text{MC/Co}$ are created from the original $\text{WC/Co}$ interface. In Tab.~\ref{tab:Delta_gamma} we give the result for $\Delta \gamma_\text{MC}$ for V, Ti, and Ta doped materials. The result is obtained from density functional theory (DFT) calculations and some computational details are given in \ref{app:comp_details}. In the calculations the bulk phases that are joined together are strained into coherency. The corresponding strain energy is cancelled, so that $\gamma_\text{WC/Co}$, $\gamma_\text{WC/MC}$, and $\gamma_\text{MC/Co}$ only represent the chemical contributions to the interface energy. The interface energy can then become negative and to model the true interfaces (see below) the elastic contribution has to be added. For all three dopants; V, Ti, and Ta, $\Delta \gamma_\text{MC}$ is positive and there is a substantial energy gain when the films are formed.

\begin{table}[h]
	\caption{Interface energies $\gamma_\text{WC/Co}$, $\gamma_\text{WC/MC}$, and $\gamma_\text{MC/Co}$ between WC/Co, WC/MC, and MC/Co interfaces, respectively, for M=V, Ti, and Ta in $\text{J/m}^2$. The metal terminated basal (0001) plane is used for WC/Co and the metal terminated (111) plans for the cubic VC, TiC, and TaC phases. For more details see \ref{app:comp_details}. The interfacial energy gain $\Delta \gamma_\text{MC}$ according to Eq. \ref{eq:deltaGamma} is also given. $e_\text{MC}$ is the elastic energy per layer in $\text{J/m}^2$ to bring the cubic films into epitaxy with the hexagonal WC phase and $(\Delta \gamma_\text{MC} - 2 e_\text{MC})$ is the resulting energy gain if the elastic energy from 2 layers is added.}
    \label{tab:Delta_gamma}
	\small
	\begin{tabular*}{\columnwidth}{@{\extracolsep{\fill}} r r r r r r r}
        \hline\hline
		$M$ & $\gamma_\text{WC/Co}$ & $\gamma_\text{WC/MC}$ & $\gamma_\text{MC/Co}$ & $\Delta \gamma_\text{MC}$ & $e_\text{MC}$ &  $(\Delta \gamma_\text{MC} - 2 e_\text{MC})$ \\	
        \hline	
		V & $1.12$ & $0.03$ & $-0.01$ & $1.10$ & $0.01$ & $ 1.08$ \\
        Ti & $1.12$ & $-0.33$ & $0.82$ & $0.63$ & $0.40$ & $ -0.17$ \\
		Ta & $1.12$ & $-0.48$ & $-0.18$ & $1.78$ & $1.39$ & $ -1.00$ \\
        \hline\hline
	\end{tabular*}
\end{table}

The positive values of $\Delta \gamma_\text{MC}$ are mainly due to the large interface energy $\gamma_\text{WC/Co}$ for the WC/Co interface compared with the interface energies $\gamma_\text{MC/Co}$ containing the cubic carbide phases (cf. Tab.\ref{tab:Delta_gamma}). An interface energy can be expressed in terms of surface energies $\sigma$ and the work of adhesion $W$ according to the Dupr\'{e} equation
\begin{equation}
 \gamma_\text{MC/Co} = \sigma_\text{MC} + \sigma_\text{Co} - W_\text{MC/Co}
\end{equation}
In Tab.\ref{tab:Dupre} the data both for the WC/Co interface with a basal (0001) hexagonal WC plane and for MC/Co interfaces (M=V, Ti, or Ta) with a cubic (111) plane are presented. Some computational details are given in \ref{app:comp_details}. The work of adhesion is a direct measure of the interface bond strength. Both for the hexagonal carbide WC and the cubic carbides MC (M=V, Ti, or Ta) the work of adhesion is similar, the interface bond strengths are similar. However, the surface energy for hexagonal WC is quite large and following the Dupr\'{e} equation the large WC surface energy $\sigma_\text{WC}$ implies a large WC/Co interface energy $\gamma_\text{WC/Co}$ (and vice versa).

The computed interface energies $\gamma_\text{WC/Co}$, $\gamma_\text{WC/MC}$ and $\gamma_\text{MC/Co}$ in Tab.\ref{tab:Delta_gamma} only contain the chemical part. To make the results more realistic the elastic contribution has to be added. In Tab.\ref{tab:Delta_gamma} we give the computed values for the elastic energy per layer $e_\text{MC}$ needed to bring the cubic films into epitaxy with the hexagonal WC phase. The mismatch varies considerably. It is $0.7\%$, $5.0\%$ and $8.2\%$ for VC, TiC and TaC, respectively, and hence the elastic energy also varies substantially. In Tab.\ref{tab:Delta_gamma} we have explicitly added the effect of two cubic layers in the last column. The interface energy gain ($\Delta \gamma_\text{MC} - 2 e_\text{MC}$) is now only positive for VC. The prediction is therefore that for V doped material epitaxial thin film formation is possible, for Ti doped material maybe, but most likely not for Ta doped material \cite{JoWa11}§.

\begin{table}[h]
		\caption{Interface energies $\gamma$, surface energies $\sigma$, and work of adhesion $W$, all in $\text{J/m}^2$, for hexagonal WC and cubic VC, TiC, and TaC. $\sigma_\text{WC}$ is the surface energy for the metal terminated basal (0001) plane, $\sigma_{MC}$ the surface energy for the metal terminated (111) surface of cubic MC (M=V, Ti, and Ta), and $\sigma_\text{Co}$ the surface energy for (111) surface of fcc Co (non-spinpolarized case). For more details see \ref{app:comp_details}.}
	\label{tab:Dupre}
	\small
	\begin{tabular*}{\columnwidth}{@{\extracolsep{\fill}} l r r r r r}
        \hline\hline
		structure & $M$ & $\gamma_\text{MC/Co}$ & $\sigma_\text{MC}$ & $\sigma_\text{Co}$ & $W_\text{MC/Co}$ \\	
        \hline	
        hexagonal & W & $1.12$ & $3.79$ & $2.45$ & $5.12$ \\
		cubic & V & $-0.01$ & $2.45$ & $2.45$ & $4.91$ \\
        cubic & Ti & $0.82$ & $3.55$ & $2.45$ & $5.19$ \\
		cubic & Ta & $-0.18$ & $2.58$ & $2.45$ & $5.21$ \\
        \hline\hline
	\end{tabular*}
\end{table}

\subsection{Layer-by-layer approach}
\label{subsec:layer_by_layer}

\begin{figure}[ht]
	\includegraphics[width=\columnwidth]{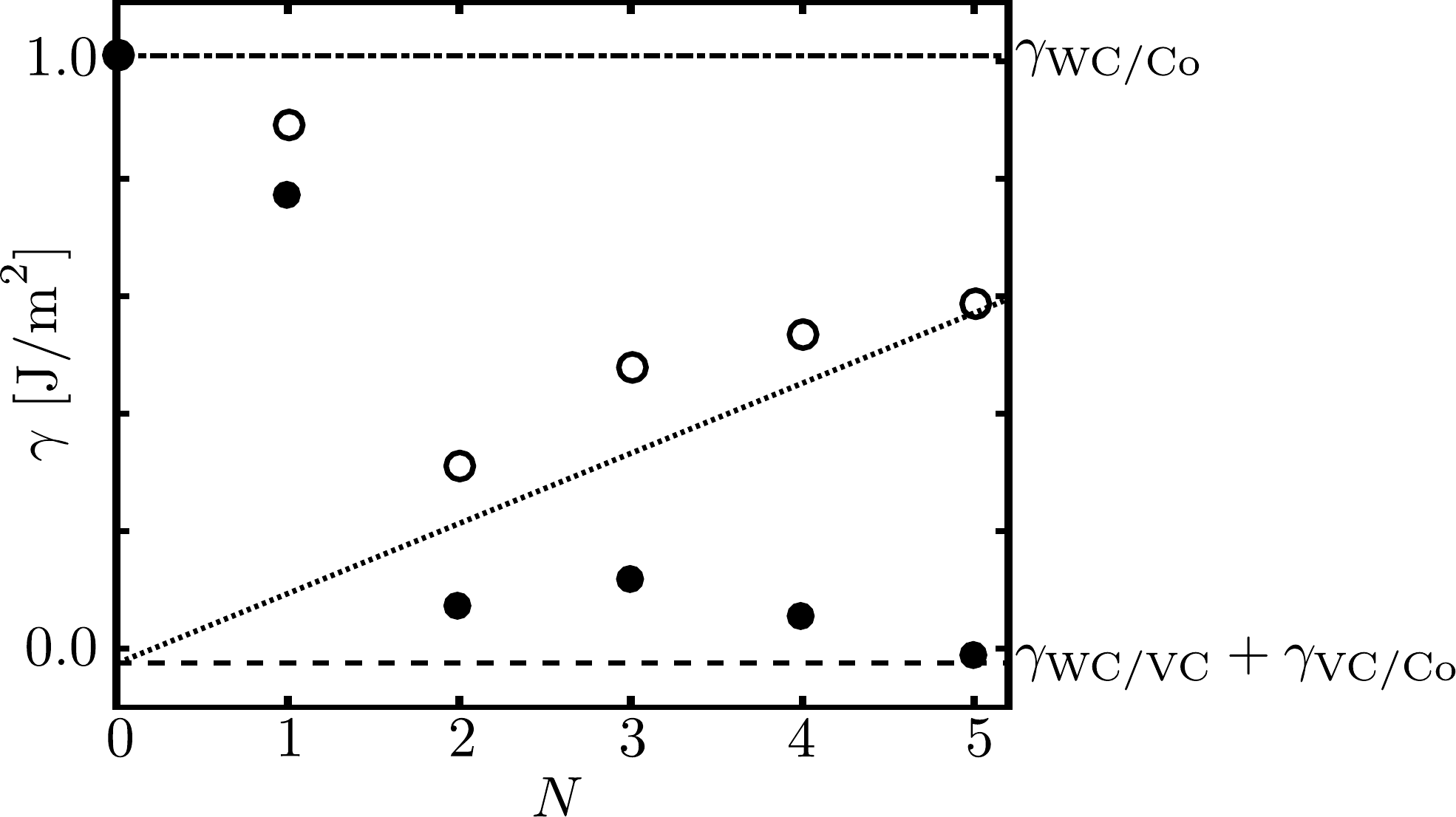}
 \caption{\label{fig:layer_by_layer}The interface energy $\gamma$ as function of number $N$ of V layers. The value $\gamma_\text{WC/Co}$=1.01 J/m$^2$
\cite{comment}
 for the clean WC/Co interface is shown as a dashed-dotted line. The filled circles show the result from the layer-by-layer calculation with no energetic cost per layer added ($\Delta g_\text{VC}+e_\text{VC}$=0). These values converge to the asymptotic value ($\gamma_\text{WC/VC} + \gamma_\text{VC/Co}$)=-0.03 J/m$^2$
\cite{comment}
 (dashed line) already after addition of a few layers. The open circles show the result when the energetic cost $\Delta g_\text{VC}+e_\text{VC}$=0.12 J/m$^2$ per layer is added, a typical value for relevant doping conditions and sintering temperatures. A clear minimum for 2 layers is obtained. This minimum will be present over a quite large temperature interval. Data from Ref. \cite{JoWa10}.}
\end{figure}

The next step in the modeling is to take the atomic scale layered structure into account. The decomposition of $\gamma_\text{film}$ in Eq.~(\ref{eq:gamma_film}) is valid only if the film is sufficiently thick, so that the interior of the film is bulk-like and the interface energies are well-defined. For thin films, $\gamma_\text{film}$ will deviate from the linear dependence on $N$. An atomic-scale layer-by-layer calculation has therefore been performed in Ref. \cite{JoWa10}. In Fig.~\ref{fig:layer_by_layer} the result for $\gamma_\text{film}$ is shown for a stoichiometric cubic VC film on the WC (0001) basal plane. The VC film is stacked in the $\left \langle 111 \right \rangle_\text{VC}$ direction with alternating layers of metal and carbon atoms. All different stacking sequences and the two different terminations of the WC basal plane have been investigated and the results for the lowest energy stacking sequence are presented in Fig.~\ref{fig:layer_by_layer} as function of the number $N$ of V layers. The theoretically predicted lowest energy stacking sequence for $N > 1$ is identical to the stacking sequence suggested from a high resolution transmission electron microscopy (HRTEM) study \cite{LaHaThLo04}.

The filled circles in Fig.~\ref{fig:layer_by_layer} show the result with no energetic cost per layer added to build up the thin film ({\it i.e.} $\Delta g_\text{VC} +  e_\text{VC} = 0$). The result should converge to ($\gamma_\text{WC/VC} + \gamma_\text{VC/Co}$) for sufficiently thick films (the dashed line). We notice that the layer-by-layer result is close to the asymptotic value already after two V layers. It implies that interface energy gain $\Delta \gamma_\text{VC}$ is essentially obtained already after the addition of two layers. The relevant interactions are short ranged. Additional layers will then only add the energy cost ($\Delta g_\text{VC} +  e_\text{VC}$) per layer and we expect to obtain a minimum in energy for a thin film with two V layers.

The energetic cost $\Delta g_\text{VC}$ depends on the temperature and on the chemical composition in the binder phase. At liquid phase sintering temperatures around 1400\C\ and for relevant doping concentrations, $\Delta g_\text{VC}$ varies between 0 and about 0.2 J/m$^2$ (0 to 0.09 eV per formula unit) \cite{JoWa10}. This has been determined using the Thermo-Calc software \cite{SuJaAn85} with a database for V additions in the W-C-Co system \cite{FrMa08}. We have added ($\Delta g_\text{VC} +  e_\text{VC}$) = 0.12 J/m$^2$ per layer (0.055 eV per formula unit) to the data in Fig.~\ref{fig:layer_by_layer}. The result is shown as open circles. It is clearly seen that the thin film structure has a minimum for 2 layers of V. Furthermore, this minimum will be present over a quite large temperature range. When the temperature is increased/decreased the slope of the dotted line will increase/decrease, but 2 layers of V will be the stable interfacial structure over a large temperature interval. When cooling and $\Delta g_\text{VC}$ approaches zero the VC phase becomes stable and it can precipitate and start to grow, most likely on the thin films at WC/Co interfaces.

The two atomic layer thick VC film at the WC/Co interface is stabilized by interfacial effects and it is thermodynamically stable over a quite large temperature range. For this structure we may use the notion, an {\it interface complexion}.

\section{Interfacial phase diagram}

We now would like to construct the actual {\it interfacial phase diagram}. It should describe the equilibrium structure and composition of the WC(0001)/Co interface in vanadium doped cemented carbides as function of the intensive thermodynamic variables temperature $T$, pressure $P$ and chemical potentials $\mu_i$. For more details, we refer to Ref. \cite{JoWa12}.

The propensity of the dopant to segregate to the interface is governed by its effect on the interface energy $\gamma$. In general, $\gamma$ can be written as \cite{SuBa96}
\begin{equation}
	\gamma = \frac{1}{A} \left( G(T) - \sum_i N_i \mu_i(T) \right).
\end{equation}
Here, $G$ is the Gibbs energy for the considered interface system, which contains a planar interface of area $A$. $N_i$ is the number of atoms in the system of component $i$ with a corresponding chemical potential $\mu_i$ determined by the reservoirs with which the interface is in equilibrium. The equilibrium interfacial state is given by the minimum of $\gamma$ under fixed $\mu_i$, $T$ and $P$.

\subsection{Interface structural models}
\label{subsec:structure}

The first step is to define an interface structural model for which an expression of $G$ for the WC(0001)/Co interface can be formulated. We are then guided by both results from high-resolution transmission electron microscope (HRTEM) imaging \cite{LaHaThLa02,LaThHaTh03} and from theoretical investigations \cite{JoWa10}. The interface is assumed to have a layered geometry where each layer has well-defined atomic lattice sites. The structural model contains a set of WC layers with hexagonal stacking, mixed layers, and Co layers describing the binder phase. The mixed layers are in epitaxy with the hexagonal WC layers and consist of two metal layers and a carbon layer in between. The metal layers can have a mixed composition of V and W atoms and the corresponding site fractions are denoted $y_\text{V}$ and $y_\text{W}$. We assume that $y_\text{V} + y_\text{W} =1$, and hence, we neglect any metal vacancies and assume no Co occupancy in the mixed layers. We also neglect any carbon vacancies in the mixed region. The carbide phase is metal terminated against the binder phase and the Co binder phase may contain dissolved W, V and C atoms. However, in the present modelling we do not explicitly include these dissolved atoms in the binder phase. As usual, the hexagonal WC bulk phase is assumed to be perfectly stoichiometric.

\begin{figure}[h]
	\includegraphics[width=\columnwidth]{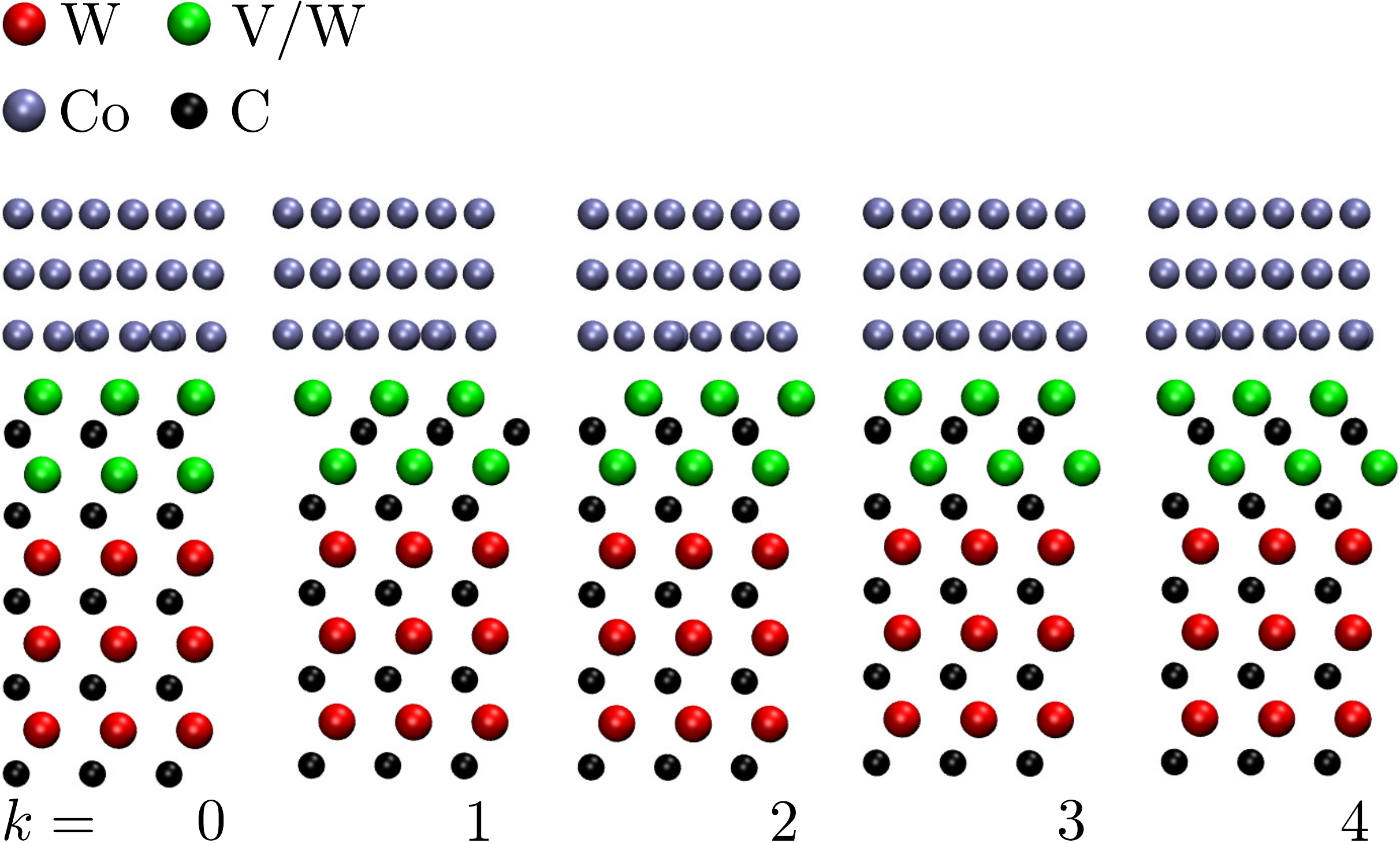}
	\caption{\label{fig:atomic_setups} The different interface structural models (labelled by $k$) viewed from the side along the $[2 \bar{1} \bar{1} 0]_\text{WC}$ direction. The lower parts show the hexagonal WC layers (red and black dots), the middle parts the mixed layers (green and black dots), and the upper parts the Co binder phase (blue dots). The plots show the relaxed atomic positions.}
\end{figure}

Several different interface structural models are constructed (cf Fig.~\ref{fig:atomic_setups}). They are denoted as $k=0,1,\ldots,4$. The model $k=0$ corresponds to hexagonal stacking in the mixed region and with $y_\text{W}=1$ ($y_\text{V}=0$) we recover the unreconstructed WC(0001)/Co interface without any segregation of V. The model $k=1$ corresponds to cubic stacking in the mixed region and with $y_\text{V}=1$ ($y_\text{W}=0$) the model derived in the layer-by-layer approach \cite{JoWa10} (see subsection \ref{subsec:layer_by_layer}) is obtained.

For each structural model $k$ there is a corresponding equilibrium interfacial state with energy $\gamma^k$. The interface energy depends on the composition in the mixed region $\gamma^k(y_\text{V},y_\text{W})$
and the equilibrium energy $\gamma^k$ is obtained by minimizing with respect to that composition. The final equilibrium interface energy $\gamma^\text{eq}$ and corresponding equilibrium structure are then given by the minimization with respect to $k$
\begin{equation}
	\label{eq:gamma_min}
	\gamma^\text{eq} = \min_k \gamma^k
    = \min_k \left[ \min_{y_\text{V},y_\text{W}} \gamma^k (y_\text{V},y_\text{W}) \right].
\end{equation}

\begin{figure}[h]
	\includegraphics[width=\columnwidth]{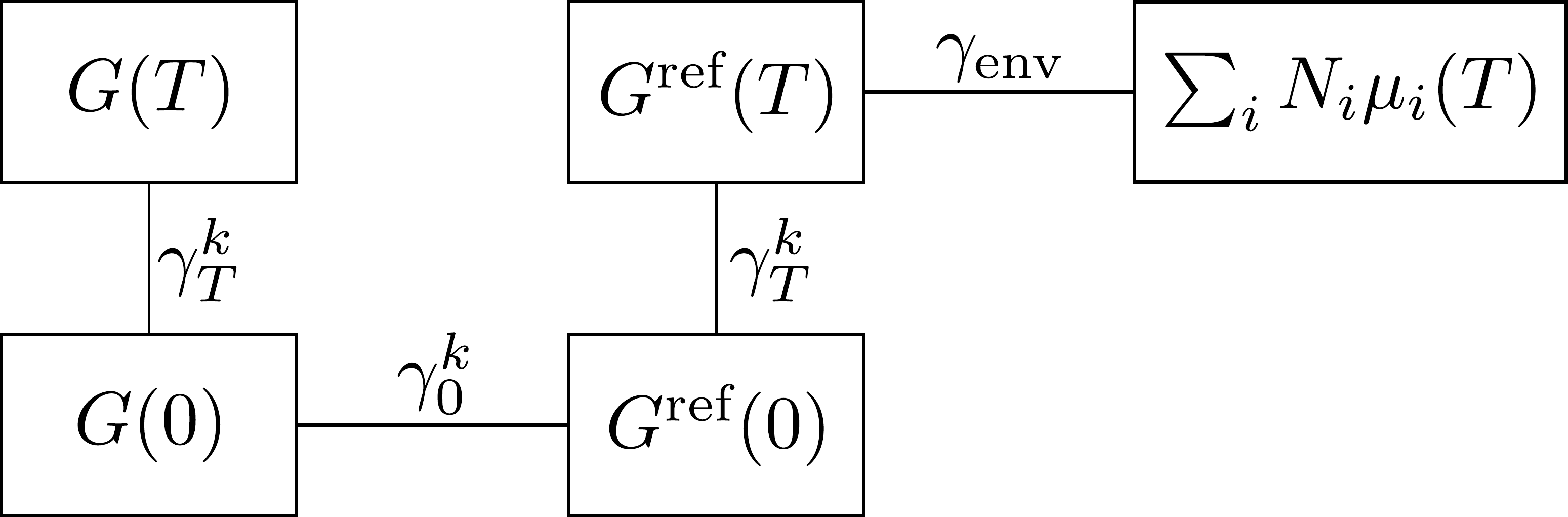}
	\caption{\label{fig:schematic} A schematic illustration of the separation of the computation of the interface energy into three different parts.}
\end{figure}

The computation of the interface energy $\gamma^k$ can be separated into three parts, a temperature independent part $\gamma_0^k$, a temperature dependent part $\gamma_T^k$ and an environment dependent part $\gamma_\text{env}$,
\begin{equation}
    \label{eq:gamma_sep}
	\gamma^k = \gamma_0^k + \gamma_T^k + \gamma_\text{env}.
\end{equation}
This is schematically illustrated in Fig.~\ref{fig:schematic}, where $G(T)$ is the Gibbs energy for the interface system and $G^\text{ref}(T)$ represents the Gibbs energy for a reference system used in the computations.

\hspace{1cm}

\subsection{Reference states}

The Gibbs energy for the reference system can be expressed as
\begin{equation}
 G^\text{ref}(T) = \sum_i N_i \mu_i^\text{ref}(T).
\end{equation}
The use of well-defined and suitable reference states with chemical potentials $\mu_i^\text{ref}$ is an important part of the modeling strategy.

In the W-C-Co system, the WC+Co two-phase field borders to the three-phase fields WC+Co+graphite and WC+Co+$\eta$ for high and low C content, respectively \cite{MaSuFr05} (where $\eta$ is a $\text{(Co,W)}_6 \text{C}$ carbide phase). We use graphite as our reference state for carbon, {\it i.e.}
\begin{equation}
	\mu_\text{C}^\text{ref}(T) = g_\text{gr}(T),
\end{equation}
where $g_\text{gr}$ is the Gibbs energy per atom of graphite. The interface system is in equilibrium with a stoichiometric WC bulk phase, hence
\begin{equation}
	\mu_\text{W}^\text{ref}(T) = g_\text{shp WC}(T) - g_\text{gr}(T),
\end{equation}
where $g_\text{shp WC}$ is the Gibbs energy per formula unit of hexagonal WC.

The amount of V added to the WC-Co system will affect $\mu_\text{V}$, and thereby the interface system. A maximum of $\mu_\text{V}$ is attained when the addition of V is so large, that a secondary carbide bulk phase forms. For V, this phase corresponds to (at typical sintering temperatures and C contents) a cubic $\text{(V,W)C}_x$ sub-stoichiometric ($x<1$) carbide of NaCl structure. To determine the vanadium reference state we use the corresponding stoichiometric carbide phase VC and define
\begin{equation}
	\mu_\text{V}^\text{ref}(T) = g_\text{NaCl VC}(T) - g_\text{gr}(T),
\end{equation}
where $g_\text{NaCl VC}$ is the Gibbs energy per formula unit of fully stoichiometric cubic VC. Finally, for cobalt we use solid fcc Co as reference state,
\begin{equation}
	\mu_\text{Co}^\text{ref}(T) = g_\text{fcc Co}(T),
\end{equation}
where $g_\text{fcc Co}$ is the Gibbs energy per formula unit of fcc Co.

\subsection{Temperature independent part}

The temperature independent part $\gamma_0^k$ in Eq.~(\ref{eq:gamma_sep}) corresponds to the thermodynamic ground state of the interface at $T=0$~K, {\it i.e.} the lowest energy configuration for given $n_\text{W}$, $n_\text{V}$, etc. For this part straight-forward DFT calculations are used to compute the energy difference
\begin{equation}
 \gamma_0^k = \frac{1}{A} \left( G(0)- \sum_i N_i \mu_i^\text{ref}(0) \right).
\end{equation}
within the methodology presented in earlier papers \cite{ChWaAlLa05,ChWaLaAl07,JoWa10,JoWa11}.

\subsection{Temperature dependent part}

The temperature dependent part $\gamma_T^k$ in Eq.~(\ref{eq:gamma_sep}) can be written as
\begin{eqnarray}
 \gamma_T^k &=& \frac{1}{A} \left( \Delta G(T) - \Delta G^\text{ref}(T) \right) \\
            &=& \frac{1}{A} \left( [\ G(T)- G(0)\ ]
 - \sum_i N_i [\ \mu_i^\text{ref}(T) - \mu_i^\text{ref}(0)\ ]\right). \nonumber
\end{eqnarray}
and it is equal to the difference of temperature dependence of the Gibbs energy between the interface system $\Delta G(T) = [G(T)- G(0)]$ and the chosen reference system $\Delta G^\text{ref}(T) = \sum_i N_i [\ \mu_i^\text{ref}(T) - \mu_i^\text{ref}(0)]$. The temperature dependence of Gibbs energy can be caused by configurational changes, vibrational motion, magnetic and electronic effects as well as melting of the binder phase.

The configurational changes in the solid phases are only present in the interface system, not in the reference system, and they are due to varying degree of W-V ordering in the mixed lattice sites. This effect is important and has been thoroughly taken into account. A cluster expansion (CE) has been developed for the mixed layers based on DFT calculated energies for a large set of different V-W configurations. The ATAT package \cite{WaCe02,WaAs02,WaAsCe02,Wa09} is used to derive the parameters in the CE by fitting to the DFT data. The CE coefficients are allowed to be different in the two different mixed layers. Monte Carlo (MC) simulations are then performed within the canonical ensemble and both the contribution to the temperature dependence of the energy (enthalpy) and entropy are determined \cite{JoWa12}.

Another contribution to the temperature dependence of the Gibbs energy is the vibrational motion. The contribution to $\gamma_T^k$ will depend on the difference in the vibrational motion for the atoms in the interface system compared with the atoms in the reference system. Due to the similarities between the interface system and the corresponding reference system we expect a large degree of cancellation to be present. In Ref. \cite{JoWa10} the vibrational contribution to $\gamma_T^k$ was estimated to be at maximum $\pm$0.3~J/m$^2$. It can be obtained from phonon spectrum calculations but the computational effort is substantial. In the present study we have therefore chosen to neglect the vibrational contribution to $\gamma_T^k$.

More complex is the deviation that arises from the temperature dependent magnetic structure of Co. From theoretical calculations, it is known that the magnetic state of Co influences Co/carbide interface energies \cite{Du02}. The absolute value of a Co/carbide interface energy can change substantially ($\sim0.4 \, \text{J/m}^2$) when treating Co as either ferro- or nonmagnetic \cite{Du02,JoWa11}, although the relative differences between various Co/carbide interfaces do not \cite{JoWa11}. Another complication is posed by the melting of Co that appear in the temperature interval relevant for sintering.

Accurate absolute values of WC/Co interface energies are therefore difficult to obtain, in particular at high temperatures. However, both the effect of Co magnetism and the liquid phase of Co should to large degree cancel when considering a difference $\gamma^{k_2}(y_{\text{V}_2}, y_{\text{W}_2}) - \gamma^{k_1}(y_{\text{V}_1}, y_{\text{W}_1})$ between interfaces of different atomic setups and site fractions calculated at the same thermodynamic conditions $T$ and $\mu_i$. Therefore, in the following, we will concentrate on the relative interface energies
\begin{equation}
	\label{eq:delta_gamma}
	\Delta \gamma^{k} (y_\text{V},y_\text{W}) = \gamma^{k} (y_\text{V},y_\text{W}) - \gamma^{k=0} (0, 1),
\end{equation}
{\it i.e.} the difference in energy between an interface $k$ containing V at site fraction $y_\text{V}$ and the unreconstructed ($k=0$) interface containing no segregation of V. Strictly speaking, this means that we are calculating \textit{free energy differences} between systems \textit{containing an interface} rather than the actual interface energy.

Furthermore, we do not consider any segregation of Co atoms into the mixed region. Neither do we consider W or V segregation on the binder phase side of the interface. Consequently, the number of Co atoms is the same in the two systems $\gamma^{k} (y_\text{V},y_\text{W})$ and $\gamma^{k=0} (0, 1)$, and hence the value for $\Delta \gamma^{k} (y_\text{V},y_\text{W})$ in Eq.~(\ref{eq:delta_gamma}) will be independent on the chemical potential for Co. It implies that the particular choice of $\mu_\text{Co}(T)$ for the current results for $\Delta \gamma$ is of no importance.

\subsection{Environment dependent part}

The environment dependent part $\gamma_\text{env}$ of Eq.~(\ref{eq:gamma_sep}) introduces the connection between the DFT computations and CALPHAD-type modeling. This is independent on the chosen structural model $k$. It is given by the difference between the chemical potentials for the chosen reference states $\mu_i^\text{ref}$ in the DFT modeling and the actual experimental chemical potentials $\mu_i$,
\begin{equation}
 \gamma_\text{env} = \frac{1}{A} \left( \sum_i N_i [\ \mu_i^\text{ref}(T) - \mu_i(T)\ ] \right).
\end{equation}
Calculating chemical potentials with respect to chosen reference states is a standard procedure in thermodynamic software. For well-defined experimental conditions (temperature, pressure and composition), the difference $(\mu_i^\text{ref}(T) - \mu_i(T))$ can thus be obtained given databases for the system at hand. An example is provided in subsection~\ref{subsec:IntPhaseDiag}.

\subsection{Interfacial phase diagram}
\label{subsec:IntPhaseDiag}

The final step is the actual construction of the interfacial phase diagram. We will make predictions for a real V-doped WC-Co material, which was produced and investigated by Weidow \textit{et al.} \cite{WeNoAn09,WeAn10}. The composition of their material is given in Tab.~\ref{tab:exampleMaterial} and details regarding its manufacture can be found in Ref.~\cite{WeNoAn09}. The V addition used is rather typical for WC grain growth inhibition and a signicantly smaller WC grain size was reported for this material compared with a reference straight WC-Co \cite{WeNoAn09}.

\begin{figure}[h]
	\includegraphics[width=\columnwidth]{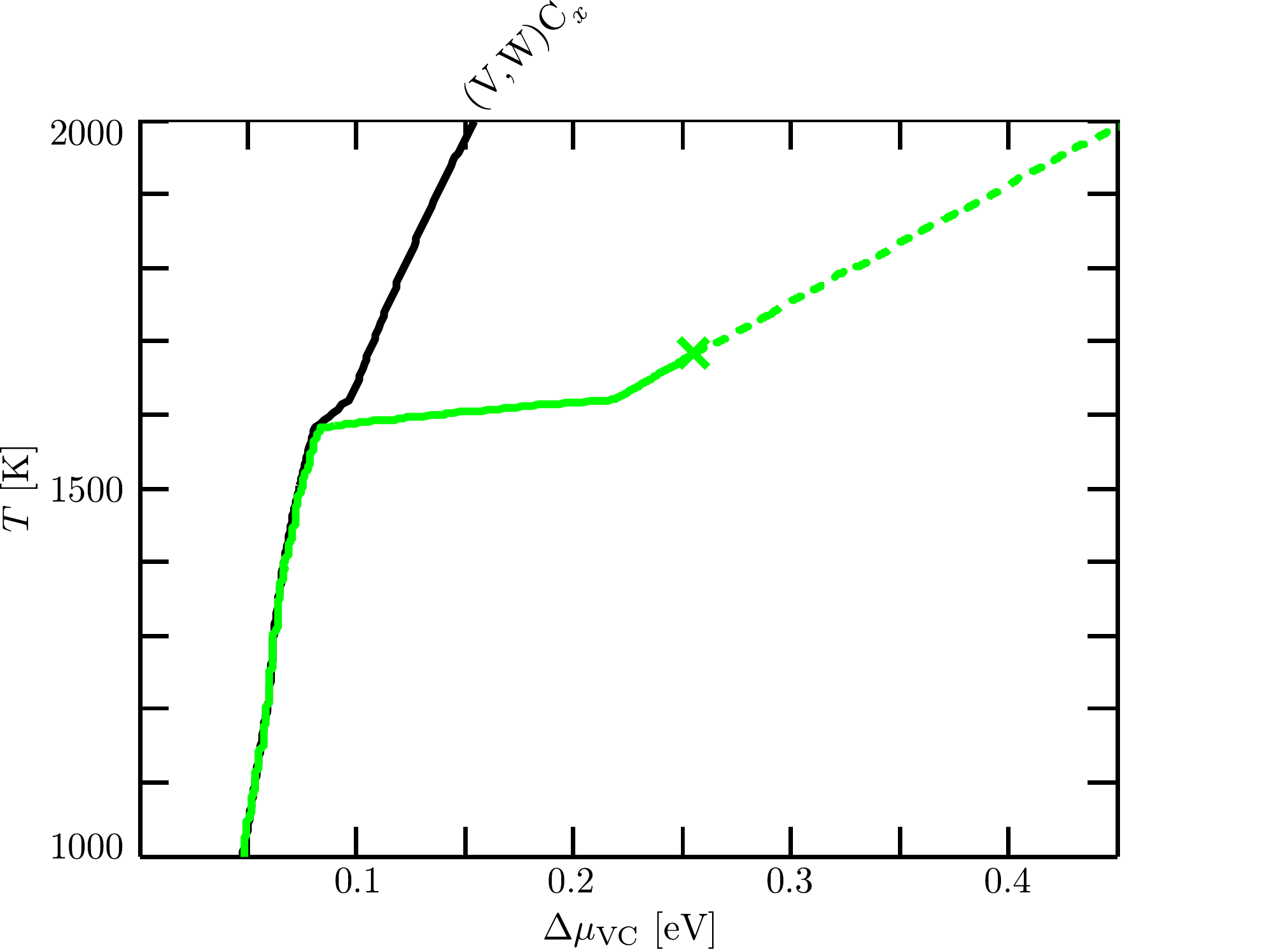}
\caption{\label{fig:PhaseDiagram}
Bulk behavior for V doped WC/Co determined using Thermo-Calc software \cite{SuJaAn85} together with data from Refs \cite{MaSuFr05,FrMa08}. The stability line of the (V,W)C$_\text{x}$ phase is shown as a black curve (see text).  To the right of the $\text{(V,W)C}_x$ stability line V is predicted to be dissolved in the binder phase. The green curve shows the calculated $\Delta\mu_\text{VC}-T$ path (cf Eq. \ref{eq:delta_mu}) that corresponds to the experimental conditions in Refs \cite{WeNoAn09,WeAn10}. Following the green curve, above 1550 K the (V,W)C$_\text{x}$ bulk phase becomes dissolved in the binder phase. The green cross marks the holding temperature during liquid phase sintering and the dashed green line indicates the behavior if one would had continued to increase the temperature in the experimental study. Data from Ref. \cite{JoWa12}.}
\end{figure}

We first determine the thermodynamic bulk behavior. For a small addition of V atoms $\mu_V$ is low and the V atoms will be dissolved in the binder phase. By increasing the amount of V $\mu_V$ will increase and a maximum will be attained when the secondary cubic $\text{(V,W)C}_x$ bulk phase is formed. This defines the $\text{(V,W)C}_x$ stability line. Using the Thermo-Calc software \cite{SuJaAn85} with appropriate databases \cite{MaSuFr05,FrMa08}, we have determined the location of the stability line. It is shown in Fig.~\ref{fig:PhaseDiagram} (black curve) as function of the difference
\begin{equation}
 \label{eq:delta_mu}
 \Delta\mu_\text{VC} = g_\text{NaCl VC} - [ \mu_\text{V} + \mu_\text{C} ]
\end{equation}
where $g_\text{NaCl VC}$ is the Gibbs energy per formula unit of the (hypothetical) stoichiometric cubic VC phase and $\mu_\text{V}$ and $\mu_\text{C}$ are the chemical potentials for V and C, respectively. The exact location of the $\text{(V,W)C}_x$ stability line in the $\Delta \mu_\text{VC}$-$T$ diagram depends also on the chemical potential for C. In Fig.~\ref{fig:PhaseDiagram} we have used the same value for $\mu_\text{C}$ as derived from the experimental set up in Refs~\cite{WeNoAn09,WeAn10}.

With increasing temperature $\Delta\mu_\text{VC}$ for the stability line increases more and more indicating that the real non-stoichiometric $\text{(V,W)C}_x$ phase becomes more and more stable compared with the (hypothetical) stoichiometric cubic VC phase. To the right of the $\text{(V,W)C}_x$ stability line V is predicted to be dissolved in the binder phase. If approaching the stability line from right the $\text{(V,W)C}_x$ bulk phase would start to nucleate and the size and thickness of such precipitates would not directly depend on thermodynamic parameters.

We can now plot in the same figure the $\Delta \mu_\text{VC}$-$T$ path that the system in the experimental study \cite{WeNoAn09,WeAn10} follows. This path is also determined from Thermo-Calc modeling and the result is shown as the green curve. Following the green curve from low temperatures the $\text{(V,W)C}_x$ bulk phase is stable up to $1550 \ \text{K}$. Above this temperature the $\text{(V,W)C}_x$ bulk phase is dissolved in the binder. At 1582 K the Co binder phase start to melt (the solidus temperature for the binder) and it is completely melted at 1620 K (the liquidus temperature for the binder). This is seen as two kinks along the $\Delta \mu_\text{VC}$-$T$ curve. Within the solidus-liquidus region, $\Delta \mu_\text{VC}$ increases sharply due to the larger V solubility in liquid Co than in solid Co. The holding temperature during liquid phase sintering is $1683 \ \text{K}$, marked by a green cross in Fig.~\ref{fig:PhaseDiagram}. If one would had continued to increase the temperature in the experimental study the system would had followed the $\Delta \mu_\text{VC}$-$T$ curve shown as a dashed green line.

\begin{figure}[h]
	\includegraphics[width=\columnwidth]{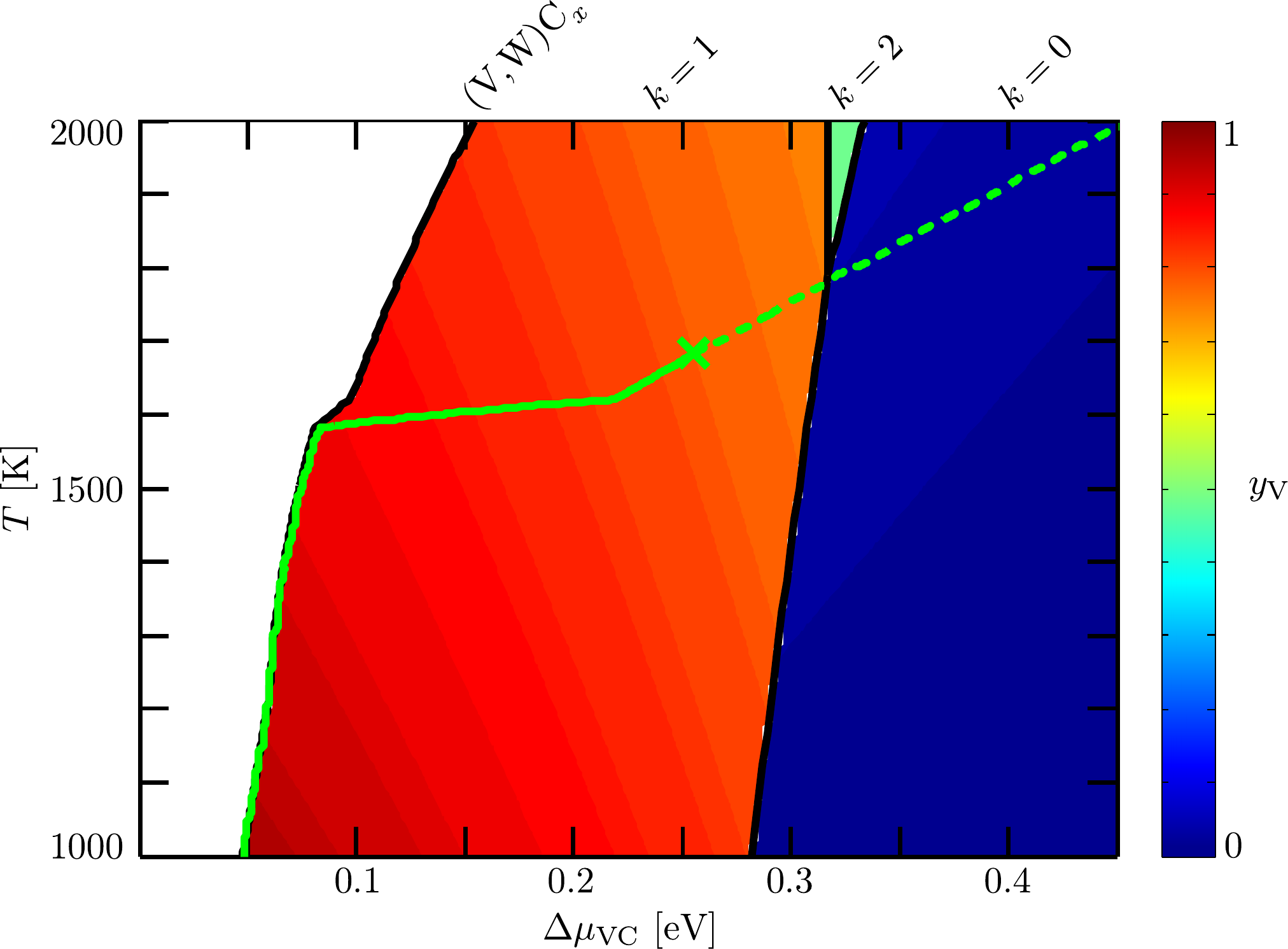}
	\caption{\label{fig:InterfacePhaseDiagram}The predicted $\text{WC}(0001)/\text{Co}$ interfacial phase diagram in V-doped WC-Co. The prediction is shown together with the results in Fig.~\ref{fig:PhaseDiagram}. The red area shows the region with high-segregation, with a two monolayer thick (V,W)C film with cubic stacking ($k=1$). The blue area corresponds to an interface with a hexagonal WC stacking ($k=0$) and with a very minor segregation of V to the topmost metal layer. The color within the regions $k=0$, $k=1$, and $k=2$, respectively, shows the equilibrium site fraction $y_\text{V}$. The black lines mark the interfacial transitions. Data from Ref. \cite{JoWa12}.}
\end{figure}

We are interested in the region to the right of the $\text{(V,W)C}_x$ stability line in Fig.~\ref{fig:PhaseDiagram} and the possibility of substantial segregation of V to interfaces and the formation of intermixed cubic structures, thin films, in that region. In Fig.~\ref{fig:InterfacePhaseDiagram} we show our results for the WC(0001)/Co interface, superimposed on the bulk result. Three different regions exist in the relevant part of the $\Delta \mu_\text{VC}$-$T$ diagram. For low $\Delta \mu_\text{VC}$, the structural model $k=1$, a thin film with cubic stacking, is associated with the lowest free energy. The two metal layers have mixed composition but are V-rich with $y_\text{V}$ ranging from $0.96$ to $0.74$ in the depicted region. Appearing only as a thin wedge in Fig.~\ref{fig:InterfacePhaseDiagram}, a region with the structural model $k=2$ expands at higher $T$ and is associated with a lower V content than $k=1$ with $y_\text{V}$ about $0.48$. At around $\Delta \mu_\text{VC}=0.3 \ \text{eV}$, the structural model $k=0$ region appears which has a low V content of, at its limits, $y_\text{V} = 0.01$ and $0.05$ at $T=1000 \ \text{K}$ and $T=2000 \ \text{K}$, respectively. This corresponds to an interface with hexagonal WC stacking with a small segregation tendency of V to the topmost layer. Thus, the three different stackings are well separated in terms of total V content; $1.5$ to $1.9$ ($k=1$), $1.0$ ($k=2$) or less than $0.1$ ($k=0$) equivalent V monolayers, and the use of the concept complexions and thermodynamic complexion transitions are well justified in the present case.

\begin{figure}[h]
	\includegraphics[width=\columnwidth]{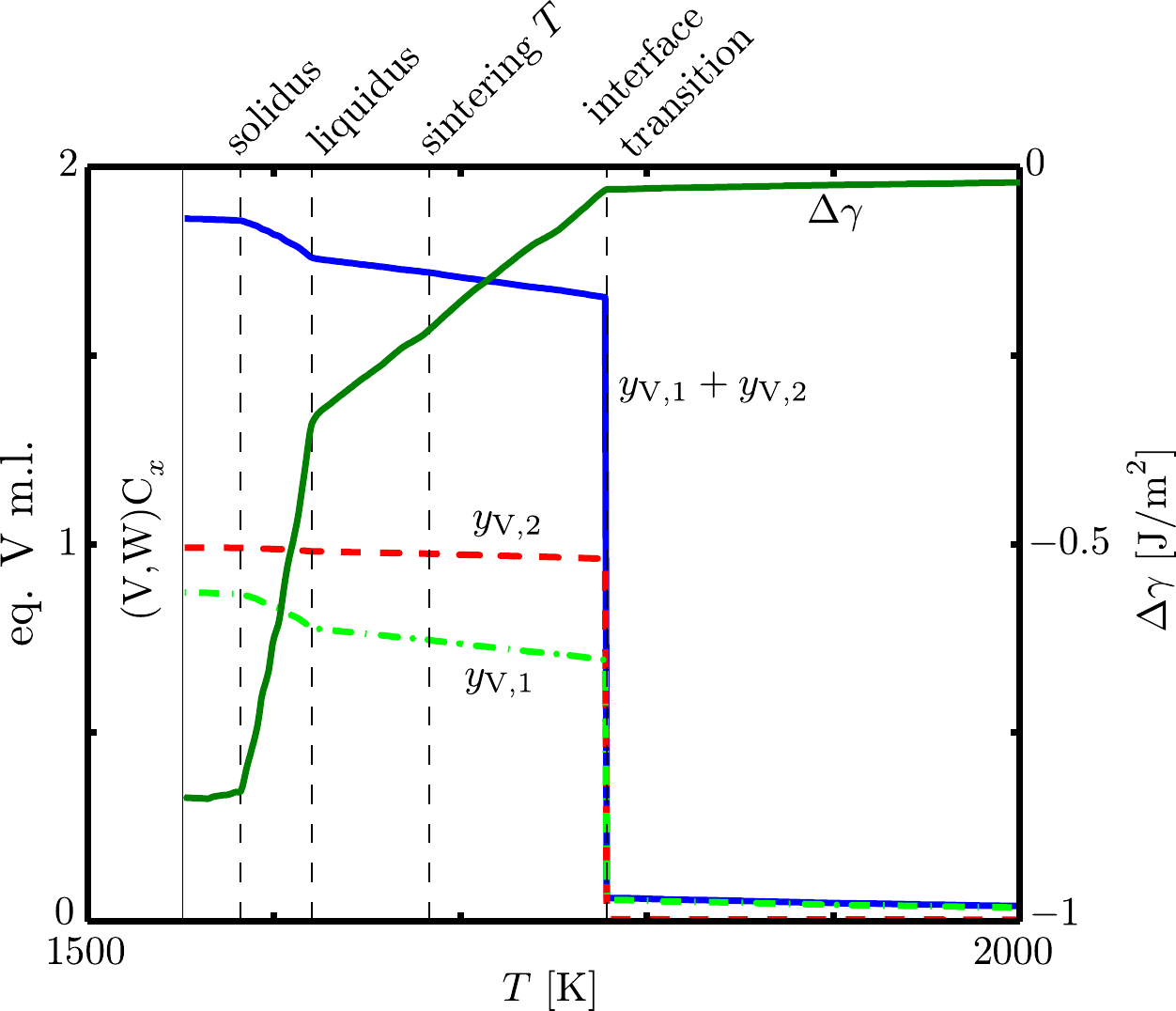}
	\caption{\label{fig:exampleMaterial}Layer-dependent site fractions $y_{\text{V},1}$ and $y_{\text{V},2}$ and total V content $y_{\text{V},1}+y_{\text{V},2}$ in terms of equivalent V monolayers (eq. V m.l.) (left axis) and relative interface energy $\Delta \gamma$ (right axis) for a $\text{WC}(0001)/\text{Co}$ interface at the experimental set up used in Refs \cite{WeNoAn09,WeAn10}. $y_{\text{V},1}$ ($y_{\text{V},2}$) is the site fraction of V in the metal atom layer closest to Co (second metal atom layer). Data from Ref. \cite{JoWa12}.}
\end{figure}

The V concentration along the experimental $\Delta \mu_\text{VC}$-$T$ path is shown in Fig.~\ref{fig:exampleMaterial} (blue curve). At $1550 \ \text{K}$ the concentration corresponds to 1.93 equivalent V monolayers (eq. V m.l.). It decreases to 1.88 eq. V m.l. when the Co binder phase has melted and it decreases further to 1.86 at $1683 \ \text{K}$, the holding temperature at the liquid-phase sintering stage. The layer-dependent site fractions are also shown as function of temperature. At $1683 \ \text{K}$, the holding temperature, the reduction of the V concentration is located exclusively in the topmost layer ($y_\text{V,1}$). If one would have continued to increase the temperature in the experimental study the theoretical prediction is that the film would have dissolved at $1780 \ \text{K}$ and above that temperature the topmost layer ($y_\text{V,1}$) would have contained only a very minor concentration of V atoms.

In the same figure we also show the temperature dependence of $\Delta\gamma$, the difference between the interface energy and the interface energy for an unreconstructed WC(0001)/Co interface (k=0) without any V segregation (cf. Eq. \ref{eq:delta_gamma}). At $1550 \ \text{K}$ the reduction is quite large, 0.84 J/m$^2$. The reduction decreases with increasing temperature and at $1683 \ \text{K}$, the holding temperature, it is only 0.14 J/m$^2$.

The important theoretical prediction from this study is that the $\text{WC}(0001)/\text{Co}$ interfaces are in the high-segregation state (k=1), they are covered by a two monolayer thick cubic (V,W)C film, throughout the liquid--phase sintering process in the experimental study by Weidow \textit{et al.} \cite{WeNoAn09,WeAn10}.

\begin{table}[h]
	\caption{The composition in atomic \% of the as-sintered material used in the experimental study in Refs \cite{WeNoAn09,WeAn10}.}
    \label{tab:exampleMaterial}
    \small
    \begin{tabular*}{\columnwidth}{@{\extracolsep{\fill}} c c c c}
        \hline\hline
		W & C & Co & V \\
        \hline
		0.4215 & 0.4167 & 0.1564 & 0.0054 \\
        \hline\hline
	\end{tabular*}
\end{table}

\section{Future challenges}

The presented interfacial phase diagram is a first step towards a complete complexion diagram for the doped WC-Co system, a technologically relevant material. A straightforward improvement of the calculation methodology would be to add an explicit calculation of the vibrational motion of the solid phases to obtain a more accurate description of the temperature dependence of the free energies. Addition of carbon vacancies in the cluster expansion would also improve the modelling and will be important when considering the effect of varying the carbon content in the WC-Co system.

More challenging is a proper description of the magnetic structure of cobalt and its liquid phase. Progress is being made in modelling paramagnetic systems \cite{KoDiGr08,StAlAb12} and for the modelling of the liquid state it may be necessary to make use of DFT-based inter-atomic potentials \cite{JuErTrNoHeNoSaAl05,PeGrWa15} and the molecular-dynamics technique to obtain sufficient sampling of relevant configurations. However, here we have focused on the calculation of free energy differences between systems containing an interface (cf Eq. \ref{eq:delta_gamma}) and rely on substantial cancelation effects. Accurate computations of the absolute value of interface energies at high temperatures for complex structures constitutes an outstanding challenge.

Our method is based on the use of defined interfacial structural models for which a quantitative expression for Gibbs energy can be formulated. Input from experiments with atomic scale structural resolution has been important to guide the construction of the interfacial structural models \cite{LaHaThLo04}. The research area would benefit from further experimental studies, in particular on the direct identification of both the atomic structure and its chemical nature of boundaries in various doped cemented carbides. Segregation in the binder phase of metal atoms and carbon to the phase boundaries should also be taken into account in future developments of the methodology. One may also envisage the use of DFT-based machine learning techniques to improve on the interfacial structural modelling.

Another key ingredient in our methodology is the use of CALPHAD-type of modelling of the adjoining bulk phases together with well-defined and appropriate reference states in the DFT calculations. The connection to real systems can then be made using available thermodynamic software \cite{SuJaAn85}. It relies on developed databases and application to other dopants and materials may require development and refinement of appropriate databases. In general, a challenge is to apply the presented methodology to other dopants and to other phase boundaries in cemented carbides. By comparing with experimental information it would then be possible to establish the validity of our approach.

Doping of cemented carbides is used to mitigate the WC grain growth in the sintering process \cite{ScBoLu95}. The phenomenon is experimentally well-established but the exact mechanism behind the grain growth inhibition is still unclear. The interfacial phase diagram can be used to tune sintering temperature and dopant addition to perform interfacial engineering. The diagrams are important as they provide information at the sintering temperature, where most of the grain growth occurs and where experimental tools with atomic scale resolution are rare \cite{SuMiTa12}. Outstanding challenges in the future are to establish the connection between the segregated interfacial structures, the {\it interface complexions}, and the growth mechanism as well as their influence on the mechanical properties of cemented carbides.

\section{Summary}

We have here reviewed a method for calculating an equilibrium interfacial phase diagram, {\it a complexion diagram}, depicting regions of stability for different interface structures as function of temperature and chemical potentials. We use DFT for interfacial energetics, Monte Carlo simulations together with cluster expansions based on DFT results for obtaining configurational free energies, and CALPHAD-type modeling for describing the thermodynamic properties of the adjoining bulk phases.

The concept {\it interface complexion} is also demonstrated using a simplified model. We then show that atomically thin VC films can be stabilized at WC/Co boundaries by interfacial effects and that two VC layers become the thermodynamically stable configuration over a wide temperature range.

\section*{Acknowledgments}

Financial support from the Swedish Research Council (VR 621-2013-5768), Sandvik and Seco Tools is gratefully acknowledged. Computations have been performed on resources provided by the Swedish National Infrastructure for Computing (SNIC) at C$^{3}$SE and NSC. Prof. Susanne Norgren and late Prof. Bo Jansson are acknowledged for providing thermodynamic data, Prof. Hans-Olof Andr\'en and Dr. Jonathan Weidow for fruitful discussions.

\appendix
\section{Computational details}
\label{app:comp_details}

All DFT calculations have been performed using the Vienna \textit{ab-initio} simulation package (VASP) \cite{KrFu96}. For the exchange-correlation effects, we employ the generalized gradient approximation (GGA). The exchange-correlation functional is approximated in the Perdew-Burke-Ernzerhof (PBE) scheme \cite{PeBuEr96} and the calculations are performed non-spinpolarized. A plane-wave implementation with projector augmented wave (PAW) potentials is used \cite{Bl96,KrJo99} and the plane-wave energy cutoff is set to 400~eV in all calculations. Partial occupancies are set with the method of Methfessel-Paxton \cite{MePa89} of first order with a smearing parameter of 0.2~eV. Atomic relaxations are performed until all atomic forces are smaller than {0.02~eV/\AA}.

For the interface energies presented in Tabs~\ref{tab:Delta_gamma} and \ref{tab:Dupre}, the modeled interface orientation is the same as in Ref.~\cite{JoWa12}. For the carbide/Co calculations, we use metal-terminated carbide slabs containing 5 M + 4 C atomic $(111)$ (for cubic MC) or $(0001)$ (for WC) layers each with 3 atoms. The Co slab contains 7 $(111)$ layers each with 4 atoms. For the carbide/Co calculations, a slab+slab setup without any vacuum region is used. For the WC/MC calculations, we use a setup with a total of 4 M, 4 W, and 7 C atomic layers. For consistency, we keep the supercell dimensions in the interface plane the same as in the MC/Co calculations, but due to incompatible stacking sequences between WC and MC, a slab+slab+vacuum ($\sim10$\ \AA\ vacuum) setup is used in the direction perpendicular to the interface plane. The values presented in Tab. \ref{tab:Delta_gamma} corresponds to the stacking sequence of lowest energy and the C chemical potential is assumed to be in the graphite limit. For the corresponding free surfaces, separate carbide and Co slabs with the same number of atoms as in the interface calculations are used. The $k$-point sampling is done with a $\Gamma$-centered grid with a division of $9\times9\times1$ to converge all interface energies to an estimated computational error within $0.03 \ \text{J/m}^2$.




\bibliographystyle{elsarticle-num}
\bibliography{CurrOp,comments}

\begin{thebibliography}{10}
\bibitem{comment}
The slight differences between the data for $\gamma_\text{WC/Co}$,
  $\gamma_\text{WC/VC}$, and $\gamma_\text{VC/Co}$ in Ref.~\cite{JoWa10} and in
  Tabs.~\ref{tab:Delta_gamma} and \ref{tab:Dupre} are explained by a difference
  in interface structure. In Ref.~\cite{JoWa10}, a mean value of energies for
  several coherent carbide/Co interfaces was taken as the carbide/Co interface
  energy, whereas here we use an incoherent carbide/Co interface with a more
  realistic Co atomic density at the interface.
\end{thebibliography}

%
%
%
\end{document}